\documentclass{pic2012}
\def\runauthor{Zuzana Feckov\'{a}}
\def\shorttitle{Azimuthally sensitive correlations}
\begin{document}

\title{Influence of momentum and charge conservation on azimuthally sensitive correlations}

\author{Zuzana Feckov\'{a}$^{1,2}$  and Boris Tom\'a\v{s}ik$^{2,3}$}

\address{$^{1}$ University of Pavol Jozef \v{S}af\'{a}rik, 
\v{S}rob\'{a}rova 2, 040 01 Ko\v{s}ice, Slovakia\\
E-mail: zuzana.feckova1@student.upjs.sk}
\address{$^{2}$ Matej Bel University,
Tajovsk\'eho 40, 97401 Bansk\'a Bystrica, Slovakia}
\address{$^{3}$ FNSPE, Czech Technical University in Prague\\ B\v{r}ehov\'a 7,
11519 Prague 1, Czech Republic}

\maketitle

\abstracts{Charge neutralisation procedure based on a Markov chain Monte Carlo algorithm is applied to a system of generated hadrons. The algorithm changes the charge of a randomly picked particle by shifting it within its isomultiplet. For baryons changes in both electric charge and baryon number are applied and the algorithm leads to charge and baryon number neutralisation. The procedure can thus be used to study the effects of the local charge and baryon number conservation. We attempt to study these together with the local momentum conservation 
and their effect on azimuthal correlator observable sensitive to local C and CP violation in 
quark-gluon plasma.}

\section{Motivation}
Recently, much attention has been paid to the suggestion that local P and CP violation can take place in the hot and dense matter produced in heavy-ion collisions. In non-central collisions, this leads to charge separation, i.e. an asymmetry in the number of positive and negative particles emitted along the angular momentum vector of the collision, the so-called Chiral Magnetic Effect \cite{Kharzeev}.

The charge separation is described phenomenlogically by adding a sine term to the Fourier decomposition of the charged-particle azimuthal distribution \cite{Voloshin}:
\begin{equation}
\frac{dN_{\alpha}}{d\Phi} = 1 + 2 v_1 \cos(\Phi_{\alpha}-\Psi_{RP}) + 2 v_2 \cos(2(\Phi_{\alpha}-\Psi_{RP})) + ... + 2 a_\alpha \sin(\Phi_{\alpha}-\Psi_{RP}) + ...,
\end{equation}

\begin{description}
\item[$\Phi_{\alpha}$, $\Psi_{RP}$]
are the azimuthal angles of the charged particle and the reaction plane, respectively ($\alpha, \beta$ standing for the charge of the particles)
\item[$v_1$, $v_2$]
are coefficients linked to the directed and elliptic flows
\item[$a_\alpha$]
describes the charge separation relative to the reaction plane.
\end{description}

Since $\langle a_\alpha \rangle = \langle a_\beta \rangle = 0$, a different observable must be employed to study the effect, using the fact that $\langle a_\alpha a_\beta \rangle \neq 0$. The observable is the azimuthal correlator $\gamma_{\alpha \beta}$, defined as follows:
\begin{eqnarray}
\gamma_{\alpha\beta} &\equiv& \langle \cos(\Phi_\alpha + \Phi_\beta - 2\Psi_{RP}) \rangle \\
&=& \langle \cos(\Phi_\alpha - \Psi_{RP}) \cos(\Phi_\beta - \Psi_{RP}) \rangle - \langle \sin(\Phi_\alpha - \Psi_{RP}) \sin(\Phi_\beta - \Psi_{RP}) \rangle \\
&=& \langle v_{1\alpha} v_{1\beta} \rangle - \langle a_\alpha a_\beta \rangle, 
\end{eqnarray}
where $\Phi_\alpha$, $\Phi_\beta$ are the azimuthal angles of the two charged particles.

It was, however, observed that the correlator $\gamma_{\alpha \beta}$ may be sensitive 
to local conservation of momentum, transverse momentum and charge. These effects have been studied separately in various publications \cite{Bzdak,Pratt}. We attempt to study the momentum, charge and baryon number conservation and their joint effect on $\gamma_{\alpha \beta}$.

\section{Our approach}

\begin{figure}[!b]
\vspace*{0.5cm}
\begin{minipage}[b]{0.48\textwidth}
\begin{center}
%\special{psfile=definicia.pdf voffset=-60 vscale=40 hscale= 40 hoffset=10 angle=0}
\centerline{\epsfxsize=1\textwidth\epsfbox{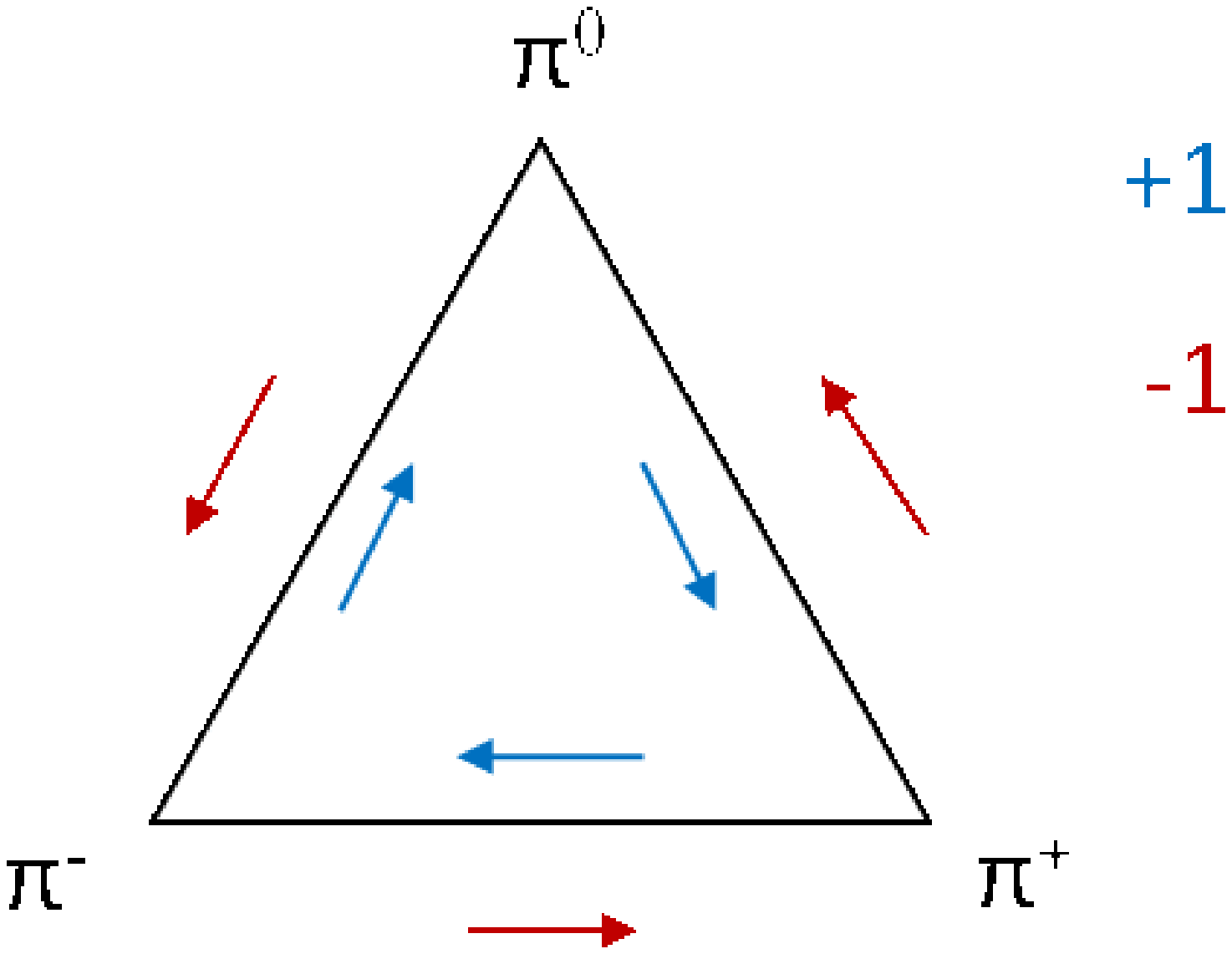}}
\caption[*]{Charge neutralisation algorithm for pions.}
\label{meson}
\end{center}
\end{minipage}
\hspace{0.03\textwidth}
\begin{minipage}[b]{0.47\textwidth}
\begin{center}
%\special{psfile=definicia.pdf voffset=-60 vscale=40 hscale= 40 hoffset=10 angle=0}
\centerline{\epsfxsize=0.88\textwidth\epsfbox{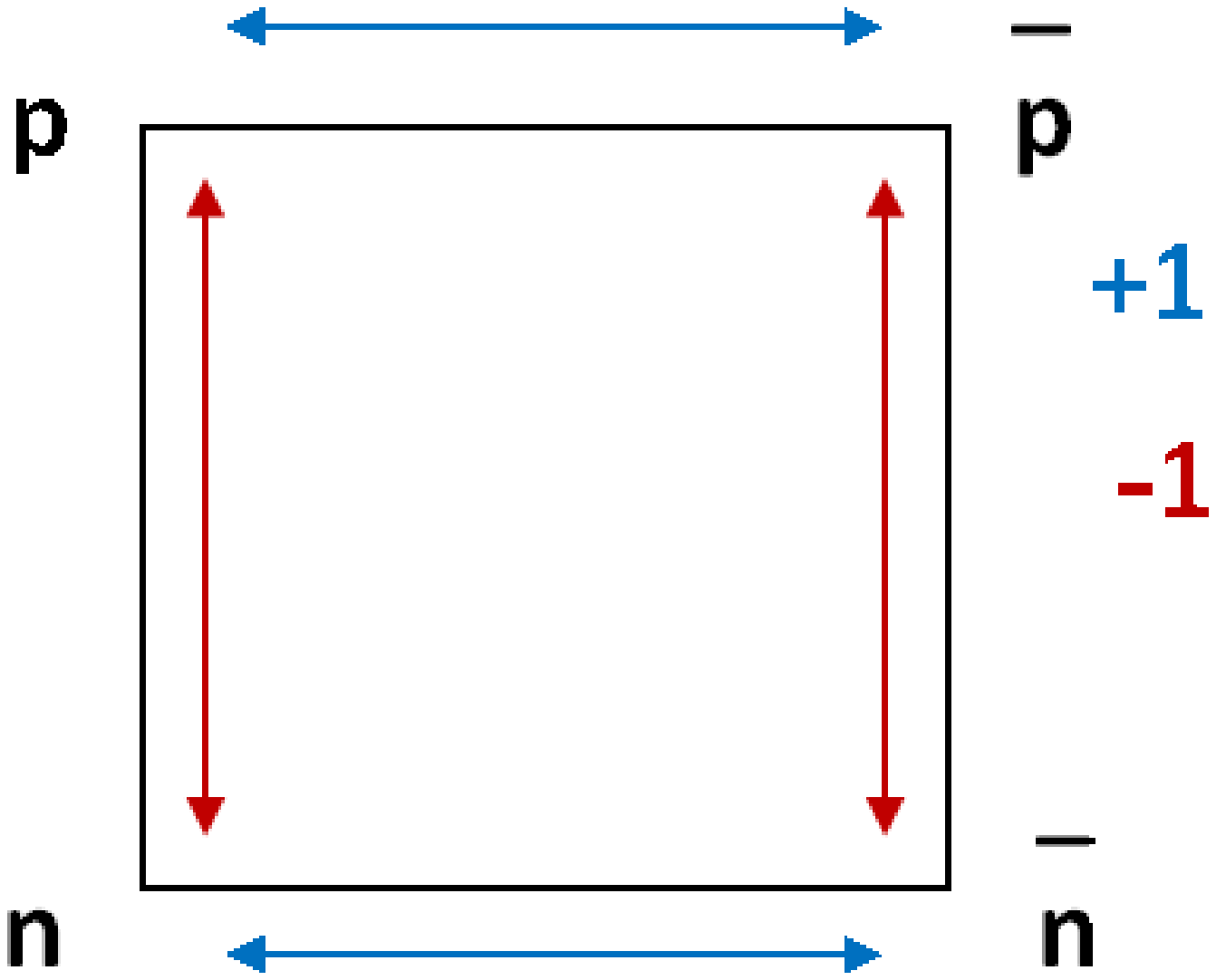}}
\caption[*]{Charge and baryon number neutralisation algorithm for nucleons.}
\label{baryon}
\end{center}
\end{minipage}
\end{figure}

The charge conservation is enforced on the system of generated hadrons by a charge neutralisation procedure based on a Markov chain Monte Carlo algorithm. The charge of a random meson is shifted within its isomultiplet. The process is illustrated in Figure~\ref{meson} 
on the pion isotriplet. The direction of the alteration is determined by another random number.

For baryons, changes in both electric charge and baryon number are applied. We show an example of the procedure in Figure~\ref{baryon} on the nucleon isodoublet. 
Either a nucleon changes into antinucleon or the proton (antiproton) and neutron (antineutron) are interchanged. The nature of the alteration is again determined by generating a random number.

\begin{figure}[!thb]
\vspace*{0.5cm}
\begin{center}
%\special{psfile=definicia.pdf voffset=-60 vscale=40 hscale= 40 hoffset=10 angle=0}
\centerline{\epsfxsize=0.6\textwidth\epsfbox{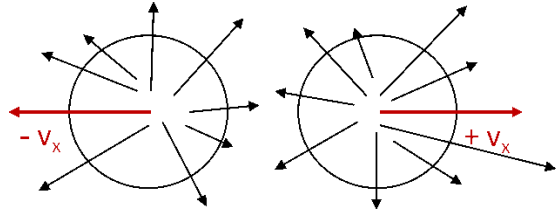}}
\caption[*]{Toy model simulation of the elliptic flow.}
\label{drops}
\end{center}
\end{figure}

The momentum conservation is ensured by generating the particles with the Lorentz Invariant Phase Space generator REGGAE \cite{Meres}. Since the azimuthal correlator γ is non-zero only in non-central collisions with elliptic flow being non-zero, we simulate the elliptic flow using two samples of hadrons moving away from each other in x-direction, as is indicated in Figure~\ref{drops}.

\section{Results}

\begin{figure}[!thb]
\vspace*{0.5cm}
\begin{center}
%\special{psfile=definicia.pdf voffset=-60 vscale=40 hscale= 40 hoffset=10 angle=0}
\centerline{\epsfxsize=2.7in\epsfbox{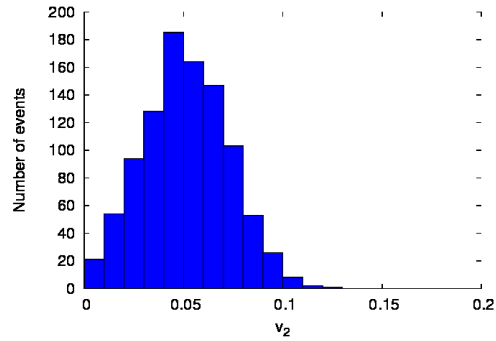}}
\caption[*]{Histogram of the simulated integrated elliptic flow $v_2$.}
\label{elliptic}
\end{center}
\end{figure}

\begin{figure}[!thb]
\vspace*{0.5cm}
\begin{center}
%\special{psfile=definicia.pdf voffset=-60 vscale=40 hscale= 40 hoffset=10 angle=0}
\centerline{\epsfxsize=4.0in\epsfbox{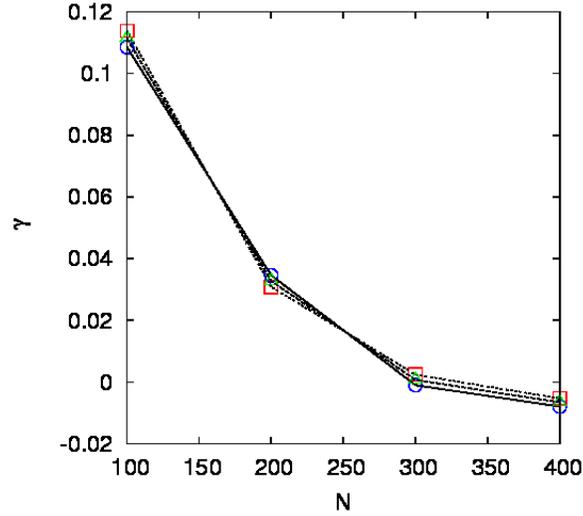}}
\caption[*]{Multiplicity dependence of the azimuthal correlator $\gamma$, triangles and dashed line correspond to pairs of particles with opposite charge, circles and solid line correspond to negative particles, squares and dotted line correspond to positive particles. }
\label{correlator}
\end{center}
\end{figure}

The variable which we specifically modelled to obtain a system similar to a non-central heavy-ion collision is the elliptic flow $v_2$. We adjusted the parameters of the simulation to find values of $v_2$ corresponding to experimental ones. The histogram of integrated $v_2$ is shown in Figure~\ref{elliptic}.

The multiplicity dependence of the azimuthal correlator $\gamma$ is shown in 
Figure~\ref{correlator}. The correlator was calculated for all pairs of charged
particles, like-sign particles (positive and negative presented separately) and unlike-sign particles. The three types are shown in Figure~\ref{correlator}. No significant difference was found between the values of the azimuthal correlator of like-sign and unlike-sign particle pairs. This induces that the contribution of joint momentum, electrical charge and baryon number conservation to the observable $\gamma$ is independent of the combined charge of the hadron pair. 

The observed signal  from the toy model qualitatively agrees with the measured correlation of unlike-sign pairs \cite{star}. We miss, however, completely the correlation of like-sign pairs. This may be due to non-inclusion of local charge conservation; charge was conserved only globally. We shall continue to investigate the topic.

\section*{Acknowledgements} 
This work was partially supported by APVV under No.~0050-11 and VEGA under 
No.~1/0457/12.

\end{document}